\documentclass[a4paper,11pt,twoside]{article}
\usepackage{booktabs}
\usepackage{amsmath,amssymb,amsfonts,amsthm,stmaryrd}
\usepackage{indentfirst}
\usepackage{color}
\usepackage{geometry}
\usepackage{tikz}
\usepackage{graphicx}
\usepackage{wrapfig}
\usepackage{fancyhdr}
\usepackage{hyperref}
\hypersetup{
    colorlinks=true,
    citecolor=blue,
    linkcolor=red,
    filecolor=magenta,
    urlcolor=black,
}

\newtheorem{theorem}{Theorem}

\newtheorem{remark}[theorem]{Remark}

\newtheorem{example}[theorem]{Example}

\let\set\mathbb

\def\exp{\operatorname{exp}}
\def\bzero{{\bf 0}}
\def\diag{\operatorname{diag}}
\def\relog{\operatorname{ReLog}}
\def\re{\operatorname{Re}}

\DeclareMathOperator{\br}{{\bf r}}
\DeclareMathOperator{\bx}{{\bf x}}
\DeclareMathOperator{\bz}{{\bf z}}
\DeclareMathOperator{\bZ}{{\bf Z}}
\DeclareMathOperator{\amoeba}{{\bf amoeba}}
\DeclareMathOperator{\crit}{{\bf crit}}

\DeclareMathOperator{\btan}{{\bf tan}}
\DeclareMathOperator{\bK}{{\bf K}}
\DeclareMathOperator{\bN}{{\bf N}}

\title{Ultimate Positivity of Diagonals of Quasi-rational Functions 
\thanks{The research was funded by the Austrian Science Fund (FWF) under
    grants Y464-N18 and W1214-N15 (project part~13)}}
\author{Hui Huang\\
  Institute for Algebra, Johannes Kepler University, Linz A-4040, Austria\\
\url{Hui.Huang@jku.at}} 
\date{}

\begin{document}
\maketitle

\section{Introduction}
The problem to decide whether a given multivariate (quasi-)rational
function has only positive coefficients in its power series expansion
has a long history. It dates back to Szeg{\" o}~\cite{Szeg1933}, who
showed that $((1-Z_1)(1-Z_2) + (1-Z_1) (1-Z_3) +
(1-Z_2)(1-Z_3))^{-\beta}$ for $\beta \geq 1/2$ is positive, in the
sense that all its series coefficients are positive, using an involved
theory of special functions. In contrast to the simplicity of the
statement, the method was surprisingly difficult. This dependency
motivated further research for positivity of (quasi-)rational
functions.  More and more (quasi-)rational functions have been proven
to be positive, and some of the proofs are even quite
simple~\cite{GRZ1983}. However, there are also others whose
positivity are still open conjectures. For instance, the rational
function $P^{-1}$ with
\[P=1-(Z_1 + Z_2 + Z_3 + Z_4) + \frac{64}{27}(Z_1Z_2Z_3 + Z_1Z_2Z_4 +
Z_1Z_3Z_4 + Z_2Z_3Z_4)\] 
is conjectured to be positive by Kauers~\cite{Kaue2007}, while no
proof is available so far. This is equivalent to verify the positivity
of the quasi-rational function $P^{-\beta}$ for $\beta \geq 1$ by
\cite[Proposition~1]{GRZ1983}. In this talk, we focus on a less
difficult but also interesting question to decide whether the diagonal
of $P^{-\beta}$ is {\em ultimately} positive, inspired
by~\cite{KaZe2008,StZu2015}. To solve this question, it suffices to
compute the asymptotics of the diagonal coefficients, which can be
done by the multivariate singularity analysis developed by Baryshnikov, 
Pemantle and Wilson~\cite{BaPe2011,PeWi2013}. Note that the ultimate positivity is a
necessary condition for the positivity, and therefore can be used to
either exclude the nonpositive cases or further support the
conjectural positivity.

\section{Multivariate singularity analysis}
In this section, we sketch and adapt the precise results for diagonals
from~\cite{BaPe2011,PeWi2013}. Reader interested in knowing more
general cases may refer to the original text.

Let $F(\bZ) = \sum_{\br\in \set N^d} a_{\br} \bZ^{\br}$ be a
$d$-variate complex generating function analytic at the origin, where
$\bZ = (Z_1,\dots,Z_d)$, $\br = (r_1, \dots, r_d)$ and $\bZ^{\br} =
Z_1^{r_1} \dots Z_d^{r_d}$. Then the {\em diagonal} of the rational
function $F$ is defined to be $\diag(F) = \sum_{n\in \set N}
a_{n,\dots,n} Z^n$.  For simplicity, we assume $F$ to be a {\em
  quasi-rational function} of the form $F(\bZ) =P(\bZ)^{-\beta}$ with
$\beta$ a real number except for nonpositive integers and $P$ a
polynomial. The zero set~$\mathcal V_P$ of $P$ in $\set C^d$ is called
the {\em singular variety} of $F$. Note that $\bzero\notin \mathcal
V_P$ since $F$ is analytic at the origin.

We aim to estimate the coefficient $a_{n,\dots,n}$
asymptotically. Similar to the univariate case, we always start with
the multivariate Cauchy integral formula
\begin{align}\label{EQ:cauchy}
  a_{n,\dots,n} = \frac1{(2\pi i)^d} \int_{T} F(\bZ)
  \frac{d{\bZ}}{\bZ^{n+1}} 
\end{align}
where $T$ is a sufficiently small torus around the origin and
$\bZ^{n+1} = Z_1^{n+1} \dots Z_d^{n+1}$.
The essential idea of the method in~\cite{BaPe2011,PeWi2013} is to
deform the contour $T$ without changing the integral (i.e.\ avoiding
the points on $\mathcal V_P$), such that the local behaviour of the
integrand at so-called {\em minimal critical points} determines the
asymptotics (under certain conditions).  To describe minimal critical
points, we need the definition of amoebas.  Following~\cite{BaPe2011},
we let
\begin{align*}
  \relog(\bZ) &= (\log|Z_1|, \dots, \log|Z_d|).
\end{align*}
Then the real set of $\relog(\bZ)$ for all $\bZ \in\mathcal V_P$ is called
the {\em amoeba} of the polynomial $P$, denoted by $\amoeba(P)$.
Note that amoebas can be computed effectively, see~\cite{Theo2002}.
By \cite[Proposition~2.2]{BaPe2011}, there exists a component~$B$ of
$\set R^d\setminus\amoeba(P)$ such that $B$ is convex and the set
$\relog^{-1} B$ is precisely the open domain of convergence of the
power series $F =\sum_{\br\in \set N^d} a_{\br} \bZ^{\br}$. Assume
that there exists a unique point ${\bx_{\min}}$ on the boundary $\partial B$
minimizing the function $-x_1-\dots-x_d$ with $\bx \in \bar B$. We call
${\bx_{\min}}$ the {\em minimizing point} for the
  diagonal. Let~$\btan_{\bx_{\min}}(B)$ denote the tangent cone to $B$
  at $\bx_{\min}$, that is,
\[\btan_{\bx_{\min}}(B) =\{{\bf b}:
{\bx_{\min}} + \epsilon {\bf b} \in B \ \text{for all sufficiently
  small}\ \epsilon >0\}.\] Let $\bN^*_{\bx_{\min}}(B)$ be the normal
cone to $\btan_{\bx_{\min}}(B)$, namely the set of vectors ${\bf v}$
such that ${\bf v}\cdot {\bf b} \leq 0$ for all ${\bf b}\in
\tan_{\bx_{\min}} B$. Then \cite[Definition~2.13]{BaPe2011} asserts
that for each $\bZ$ with $\relog(\bZ) = \bx_{\min}$ there is a naturally
defined cone $\bK(\bZ)$ (which is too lengthy to give here) that
contains $\btan_{\bx_{\min}} (B)$. We denote $\bN^*(\bZ)$ for the normal
cone of $\bK(\bZ)$ and define the set of {\em minimal critical points}
by
\[\crit(H,\bx_{\min}) = \{\bZ\in \mathcal V_P\cap \relog^{-1}(\bx): (1,\dots,1) \in \bN^*(\bZ)\}.\]
Note that $\bN^*(\bZ)\subseteq \bN^*_{\bx_{\min}}(B)$. When $P$ is
irreducible, for $\bZ$ to be a smooth minimal critical point in the
sense that the gradient of $P$ at $\bZ$ is nonzero, we must have
\begin{align}\label{EQ:criticalpointeq}
  P(\bZ) =0 \quad \text{and}\quad Z_j
  \frac{\partial}{\partial Z_j}P(\bZ) = Z_k\frac{\partial}{\partial Z_k}
  P(\bZ), \quad j,k = 1,\dots,d.
\end{align}
We are mainly interested in the following quadratic case.
\begin{theorem}[{\cite[Proposition~3.7]{BaPe2011}}]\label{THM:coneasympt}
  Let $F(\bZ) =P^{-\beta}(\bZ)$ be a $d$-variate quasi-rational
  function with $\beta \neq 0,-1,-2,\dots$ and $P$ a polynomial. Let
  $B$ be a component of $\set R^d\setminus \amoeba(P)$ so that $F$ has
  a convergent power series expension in $\relog^{-1}(B)$. Assume that
  there exists a minimizing point $\bx_{\min}$ for the diagonal, and
  the set $\crit(H,\bx_{\min})$ contains only one point
  $\bZ_*$. Further assume that the leading homogeneous part $q$ of
  $P\circ \exp$ at $\bz_*$ with $\bZ_* =
  \exp(\bz_*)=(\exp({z_*}_1),\dots,\exp({z_*}_d))$ is an irreducible
  quadratic with the matrix $M_q$ congruent to the $d\times d$ diagonal
  matrix $\diag(1,-1,\dots,-1)$. Then, when the Gamma functions in the
  denominator are finite, 
  \begin{equation}\label{EQ:asympt}
    a_{n,\dots,n} \sim \frac{((-1)^{d-1}\det(M_q))^{-1/2}}{2^{2\beta-1} \pi^{d/2-1} \Gamma(\beta)
      \Gamma(\beta+1-d/2) }\mathbf Z_*^{-n} (n^2 q^*({\bf 1}))^{\beta-d/2},
  \end{equation}
  where $q^*$ is the dual quadratic form of $q$ with the matrix
  $M_q^{-1}$. 
\end{theorem}

\section{Asymptotics of diagonals}
In this section, we apply the multivariate singularity analysis to two
quasi-rational functions. The first example comes from a well-known
rational function, which was shown to be positive for $\beta =1$ in~\cite{Szeg1933,AsGa1972}.
\begin{example}\label{EX:szego}
Consider the quasi-rational function
\[F(\bZ) = \frac{1}{(1-(Z_1+Z_2+Z_3)+\frac34(Z_1 Z_2 + Z_1 Z_3 + Z_2
  Z_3))^\beta} \quad \text{with} \ \beta \neq 0, -1,-2,\dots.\] 
We are interested in the asymptotics for the diagonal coefficents of
$F$. For simplicity, we tranlate each coordinate $Z_j$ to
$\frac23Z_j$, and then apply the method to $P^{-\beta}$ with
\[P(\bZ) = 1-\frac23(Z_1+Z_2+Z_3) + \frac13 (Z_1 Z_2 + Z_1 Z_3 + Z_2 Z_3).\]

\noindent{\bf Identify minimal critical points.} 
Let $\mathcal V_P $ be zero set of $P$. It is readily seen that
$\mathcal V_P$ is smooth except for the point $(1,1,1)$. Let $B$ be
the component of $\set R^3 \setminus\amoeba(P)$ corresponding to the
power series expension of~$P^{-\beta}$ at the origin. Then $B$
contains the negative orthant by~\cite{Theo2002}. We claim that
$(0,0,0)$ is on the boundary of $B$. Indeed, it suffices to verify
that $P$ is nonzero in the open unit polydisk $\{\bZ\in \set C^3:
|Z_j|<1\}$. Following~\cite[Section~4.4]{BaPe2011}, it is equivalent
to show that $P(\bZ+{\bf 1})$ is nonzero in the open disk $\{\bZ\in
\set C^3: |Z_j+1|<1\}$ by sending $\bZ$ to $\bZ+{\bf 1}$. Then further setting
$\bZ$ to ${\bf 1}/\bZ$ changes the problem to prove that
\[P\left({\bf 1}+\frac{\bf 1}{\bZ}\right)= \frac{Z_1 +
  Z_2 + Z_3}{3 Z_1 Z_2 Z_3}\] 
is nonzero in $ \{\bZ \in \set C^3: \re(Z_j) < -1/2\}$,
which is trivial since $\re(Z_1+Z_2+Z_3) < -3/2$. Since the diagonal
direction $(1,1,1)\in \bN^*_{\bzero}(B)$, the point $(0,0,0)$ is the
minimizing point for the diagonal by the definition of normal cones.
At the point $(1,1,1)$, the leading homogeneous term of $P$ composing
with the exponential is
\[q(\bZ)=\frac13(Z_1 Z_2 + Z_1 Z_3 + Z_2 Z_3)\]
with the matrix
congruent to the $3\times 3$ diagonal matrix $\diag(1,-1,-1)$.
Then the dual quadratic form of $q$ is
\[q^*(\br) = 
\begin{pmatrix}
r_1 & r_2 & r_3
\end{pmatrix} 
\begin{pmatrix} 
0 & \frac16 & \frac16\\
\frac16 & 0 & \frac16\\
\frac16 & \frac16 & 0
\end{pmatrix}^{-1}
\begin{pmatrix}
r_1\\
r_2\\
r_3
\end{pmatrix}
=3 (2 r_1 r_2 +2 r_1 r_3 +2 r_2 r_3-r_1^2 - r_2^2- r_3^2 ).\] By
definition, the normal cone $\bN^*({\bf 1})$ is the set $\{\br\in \set
R^3: q^*(\br)>0\}$ containing the diagonal direction $(1,1,1)$. Hence
the point $(1,1,1)\in\crit(H,\bzero)$. The remaining points in
$\crit(H,\bzero)$ could only be smooth critical points on $\mathcal
V_P\setminus \{(1,1,1)\}$. Solving~\eqref{EQ:criticalpointeq} implies
that $\crit(H,\bzero)$ has only one point, namely $(1,1,1)$. To get
the leading term of the asymptotics, it suffices to compute the
contribution from $(1,1,1)$.

\smallskip\noindent
{\bf Compute diagonal asymptotics.}
So far, we have shown that $P^{-\beta}$ satisfies the hypotheses in
Theorem~\ref{THM:coneasympt}. Since $F(\bZ) = P(\frac32 \bZ)^{-\beta}$,
by~\eqref{EQ:asympt} the $n$-th diagonal coefficient of
$F$ is asymptotic to
\begin{equation}\label{EQ:ex1}
   \frac{3^{2 \beta-\frac32}n^{2\beta-3}}{
  2^{2\beta-2}\sqrt{\pi}\Gamma(\beta)\Gamma(\beta -1/2)}\left(\frac23\right)^{-3n},
\end{equation}
which implies that the diagonal of $F$ is ultimately positive for $\beta>1/2$.

\end{example}

Now let's turn to the function that we mention in the introduction.
\begin{example}[\cite{Kaue2007}]\label{EX:kauers}
  Consider the quasi-rational function
  \[F(\bZ) = \frac{1}{(1- (Z_1 + Z_2 + Z_3 + Z_4)+\frac{64}{27}
    (Z_1Z_2Z_3 + Z_1 Z_2 Z_4+Z_1 Z_3Z_4 + Z_2 Z_3 Z_4))^\beta}\] with
  $\beta \neq 0, -1,-2,\dots$. 
  We want to know the diagonal asymptotics of $F$. Similar to the
  previous example, we first translate each coordinate $Z_j$ to
  $\frac38 Z_j$ and work with $P^{-\beta}$ where
  \[P(\bZ) =1- \frac38 (Z_1 + Z_2 + Z_3 + Z_4) + \frac18
  (Z_1Z_2Z_3 + Z_1 Z_2 Z_4+Z_1 Z_3Z_4 + Z_2 Z_3 Z_4).\] 
  Then the diagonal asymptotics of $F(\bZ) = P(\frac83\bZ)^{-\beta}$
  can be easily computed.

\smallskip \noindent
{\bf Identify minimal critical points.} 
Let $\mathcal V_P$ be the zero set of $P$. Then the only non-smooth
point on $\mathcal V_P$ is the point $(1,1,1,1)$. Again the component
$B$ of $\set R^4\setminus\amoeba(P)$ corresponding to the power series
expansion of~$P^{-\beta}$ at the origin is the one contains the
negative orthant. Again, we claim that $(0,0,0,0)$ is on the boundary of
$B$. Similarly, it suffices to verify that $P$ is nonzero in the open
unit polydisk $\{\bZ\in \set C^4: |Z_j| < 1\}$, which is then equivalent
to show that the numerator of
\[P\left({\bf 1}+\frac{\bf 1}{\bZ}\right)= \frac{Z_1 + Z_2 + Z_3 + Z_4
  + 2 (Z_1 Z_2 + Z_1 Z_3 + Z_1 Z_4 + Z_2 Z_3 + Z_2 Z_4 + Z_3 Z_4)}{8
  Z_1 Z_2 Z_3 Z_4}\] 
is nonzero in $D=\{\bZ\in \set C^4: \re(Z_j)<-1/2\}$. This can be done
by cylindrical algebraic decomposition (CAD)~\cite{Coll1975,Kaue2011} as
follows. Suppose that $(Z_1,Z_2,Z_3,Z_4)\in D$ is a zero of the
numerator. Then we can represent $Z_4$ as
\[ Z_4 = \frac{-Z_1 - Z_2 - Z_3 - 2 Z_1 Z_2 - 2 Z_1 Z_3 - 2 Z_2
  Z_3}{1+2 Z_1 + 2 Z_2 + 2 Z_3},\] 
whose denominator cannot be zero since $\re(Z_j)<-1/2$ for $j=1,2,3$.
Applying CAD shows that the real part of $Z_4$, in terms of real
and imaginary parts of $Z_1,Z_2,Z_3$, must be greater than $-1/2$, a
contridiction. Since the diagonal direction $(1,1,1)$ belongs to
$\bN^*_{\bzero}(B)$, the point $(0,0,0,0)$ is the minimizing point for
the diagonal by the definition of normal cones. At the point
$(1,1,1,1)$, the leading homogeneous term of $P\circ \exp$ is
\[q(\bZ) = \frac14( Z_1 Z_2 + Z_1 Z_3 + Z_1 Z_4 + Z_2 Z_3 + Z_2 Z_4 + Z_3 Z_4)\]
with the matrix
\[
\begin{pmatrix} 
0 & \frac1{8} & \frac1{8} & \frac1{8}\\
\frac1{8} & 0 & \frac1{8} & \frac1{8}\\
\frac1{8} & \frac1{8} & 0 & \frac1{8}\\
\frac1{8} & \frac1{8} & \frac1{8} & 0
\end{pmatrix}
\]
congruent to the $4\times 4$ diagonal matrix $\diag(1,-1,-1,-1)$.
Then the dual quadratic form is
\[q^*(\br) =\frac{16}{3} (r_1 r_2 + r_1 r_3 + r_1 r_4 +
r_2 r_3 + r_2 r_4+ r_3 r_4-r_1^2 - r_2^2 - r_3^2 - r_4^2).\]
Then the normal cone $\mathbf N^*({\bf 1})= \{\br\in
\set R^4: q^*(\br)>0\}.$ By the same reason as the
previous example, the set $\crit(H,\bzero)$ only contains the
point $(1,1,1,1)$, which determines the leading term of the
asymptotics.

\smallskip\noindent
{\bf Computing diagonal asymptotics.}
We have seen that $P^{-\beta}$ satisfies the hypotheses in
Theorem~\ref{THM:coneasympt}. Hence by~\eqref{EQ:asympt}, the $n$-th
diagonal coefficient of $F$ is asymptotic to
\[\frac{2^{3\beta-3} n^{2\beta-4}}{3^{\beta-\frac32}\pi \Gamma(\beta)\Gamma(\beta-1)} \left(\frac38\right)^{-4n},\]
which implies that the diagonal of $F$ is ultimately positive for
$\beta>1$.
\end{example}

\begin{remark}
  For the cases when the Gamma functions in the denominator
  of~\eqref{EQ:asympt} is infinite, e.g., when $\beta = \frac12$ for
  Example~\ref{EX:szego} and $\beta = 1$ for Example~\ref{EX:kauers},
  we need more techniques from~\cite{PeWi2013} to compute the diagonal
  asymptotics, which will be addressed in future work.
\end{remark}

\section{Acknowledgement}
I would like to thank my advisor Manuel Kauers for encouraging me to
work with this topic and also providing valuable comments.

\def\cprime{$'$} \def\cprime{$'$} \def\cprime{$'$} \def\cprime{$'$}
  \def\cprime{$'$} \def\cprime{$'$} \def\cprime{$'$} \def\cprime{$'$}
  \def\polhk#1{\setbox0=\hbox{#1}{\ooalign{\hidewidth
  \lower1.5ex\hbox{`}\hidewidth\crcr\unhbox0}}} \def\cprime{$'$}

\end{document}